\documentclass[useAMS,usenatbib,twocolumn,preprintnumbers,footinbib]{revtex4}
\usepackage{epsfig}
\input psfig.sty

\usepackage{graphicx}
\usepackage{dcolumn}
\usepackage{bm}
\usepackage{epsfig}
\usepackage{graphicx}
\usepackage{dcolumn}
\usepackage{bm}
\usepackage[english]{babel}

\begin{document}

\title{The robustness of angular diameter distances to the lens in the B1608+656 and RXJ1131-1231 systems }

\author{R. F. L. Holanda$^{1,2,3}$\footnote{E-mail: holanda@uepb.edu.br}}

\affiliation{$^1$Departamento de F\'{\i}sica, Universidade Estadual da Para\'{\i}ba, 58429-500, Campina Grande - PB, Brasil}

\affiliation{$^2$Departamento de F\'{\i}sica, Universidade Federal de Campina Grande, 58429-900, Campina Grande - PB, Brasil}

\affiliation{$^3$Departamento de F\'{\i}sica, Universidade Federal do Rio Grande do Norte, 59078-970, Natal - RN, Brasil}

\date{\today}

\begin{abstract}

The angular diameter distance of lens, $D_{Aol}$, of strong gravitational lensing systems has been claimed as a cosmic standard ruler. The first measurements for this distance were recently obtained to two well-known systems: B1608+656 and RXJ1131-1231. However, there is a range of possible systematic uncertainties which must be addressed in order to turn these systems into useful cosmic probes. In this paper, we confront $D_{Aol}$  with luminosity distances of type Ia supernovae  and angular diameter distances of galaxy clusters to search for tensions between these cosmological measurements using the cosmic distance duality relation. No tension was verified with the present data, showing the robustness of the assumptions used to describe the lens systems.

\end{abstract}


\maketitle

\section{Introduction}

Gravitational lensing is a important effect arising from Einstein's theory of general relativity, where, in simple words,  mass bends light. The most extreme bending of light occurs in the so-called strong gravitational lenses, which happens when a very massive object act as lens of an aligned source object. In this case more than one image of the source will be detected by the observer (a complete discussion on this effect can be found in Schneider, Ehlers \& Falco 1992 and Schneider, Kochanek \& Wambsganss 2006). Nowadays, this phenomenon has played a very important role in the fields of cosmology and astrophysics, probing the nature of dark matter in the universe as well as testing cosmological models (Zhu 2000; Chae 2003; Chae et al. 2004; Mitchell et al. 2005; Zhu \& Mauro 2008a; Zhu et al. 2008; Biesiada, Malec \& Pi´orkowska 2010; Yuan \& Wang 2015; Cao et al. 2015; Linder 2016).
\begin{figure*}
\centering
\includegraphics[width=0.45\textwidth]{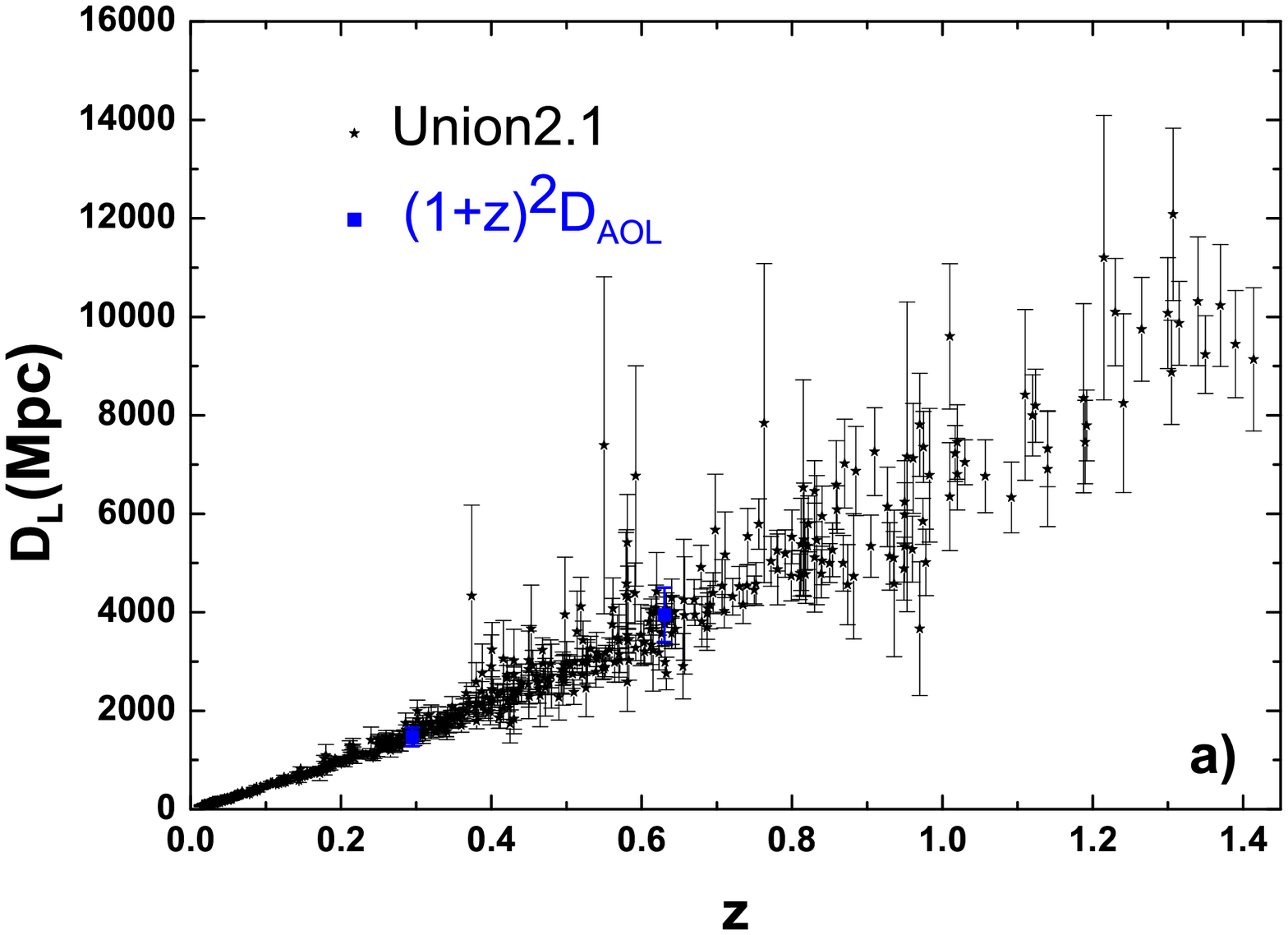}
\includegraphics[width=0.45\textwidth]{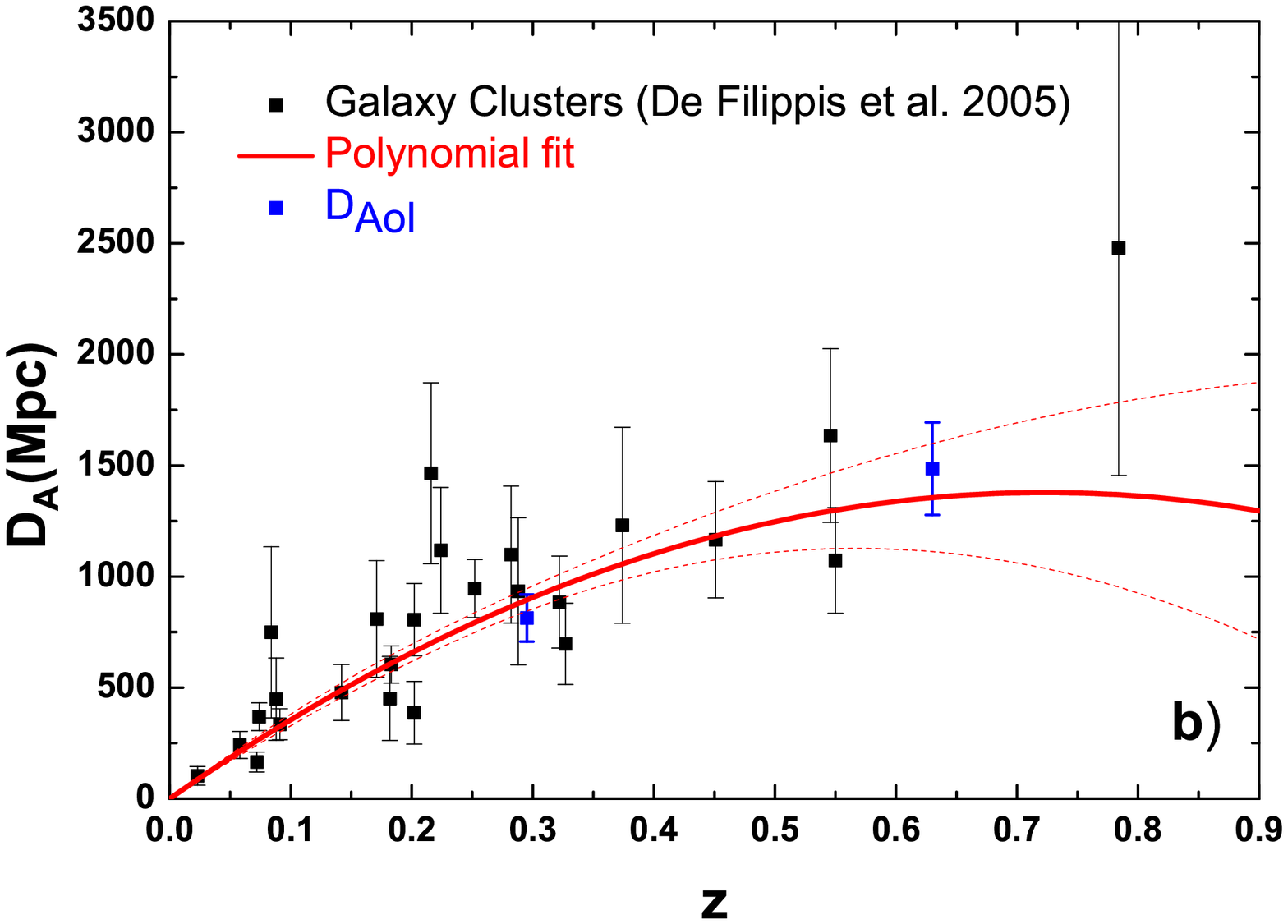}

\caption{ In fig. (a), the blue squares and black stars are, respectively, the $D_{Aol}$'s to the B1608+656 and RXJ1131-1231 systems and the luminosity distances of the SNe Ia from Union 2.1 (Suzuki et al. 2012). In fig. (b), the blue squares and black squares are the $D_{Aol}$'s to the B1608+656 and RXJ1131-1231 systems and  25  ADD of galaxy clusters (De Filippis et al. 2005), respectively. The red solid and dashed lines are the polynomial fit and  the 1$\sigma$ error, respectively.}
\end{figure*}

 Cosmological parameters can be inferred from strong gravitational lensing observation through different quantities. A known quantity is the Einstein radius, which varies with cosmological models via the ratio of angular diameter distances (ADD) between lens/source and observer/source. It occurs when the source (s), the lens (l) and the observer (o) are so well aligned that the observer source direction (Schneider, Kochanek \& Wambsganss 2006). Another one is the so-called time-delay measurement, $\Delta t$, caused by the difference in length of the optical paths and the gravitational time dilation for the ray passing through the effective gravitational potential of the lens  (Refsdal 1964; Schneider et al. 1992; Schneider et al. 2006; Treu 2010; Suyu et al. 2010). This quantity depends on the angular diameter distances between observer and lens, observer and source, and lens and source, such as (Paraficz \& Hjorth 2009; Suyu et al. 2010)  

\begin{equation}
  \Delta t = \frac{1+z_{l}}{c}\frac{D_{Aos}D_{Aol}}{D_{Als}}
  \left( \frac{1}{2} (\vec{\theta}-\vec{\beta})^2-\Psi(\vec{\theta})\right),
  \end{equation}
where $\vec{\theta}$ and $\vec{\beta}$ are the positions of the images and the source respectively, $z_{l}$ is the lens redshift, and $\Psi$ is the effective gravitational potential of the lens. The quantity $(1+z_{l})\frac{D_{Aos}D_{Aol}}{D_{Als}}$ is so-called time-delay distance, $D_{A \Delta t}$. These measurements, $\Delta t$ or $D_{A \Delta t}$, are  more sensitive to the Hubble constant, $H_0$, than other cosmological quantities since each of the ADD is proportional to the inverse of $H_0$ (Coe \& Moustakas 2009; Linder 2011; Jackson 2015).

As a new approach, Paraficz \& Hjorth (2009), by using simulated data,  showed that a cosmic standard ruler can be constructed from the joint measurement of the time-delay  between gravitationally lensed quasar images and the velocity dispersion ($\sigma^2$) of the lensing galaxy. In other words, these authors showed that one can obtain the ADD between observer and lens, $D_{Aol}$, from the ratio $\Delta t/\sigma^2$, getting more cosmological information from the time-delay lenses (see next section). The main advantage of this technique is being independent of the source redshift. In this line, by using a mock sample, Jee et al. (2015) showed that the combination of the classical time-delay distance, $D_{\Delta t}$, with the $D_{Aol}$ breaks the degeneracy between the curvature of the Universe and the time-varying equation of state of dark energy, improving  significantly the cosmological constraints from lens observations. 

On the other hand, the uncertainty in $D_{Aol}$ is expected to be dominated by the velocity dispersion uncertainty and there is a range of possible systematic uncertainties which  must be addressed when the lens system is modeled by an singular isothermal sphere (SIS)  density profile, {such as: velocity anisotropy, the fact of dynamical masses obtained under the assumption that mass follows light do not match with the masses of strong gravitational lens systems of similar velocity dispersions (Barnabe \&  Koopmans 2007; Barnabe, Spiniello \& Koopmans 2014), total mass profile shape (Schwab et al. 2009), the detailed density structures of the lenses (Tonry 1983),  mass along the line of sight to the Quasi-Stellar Objects (QSO) (Lieu 2008),  the mass-sheet degeneracy (MSD) (Falco, Gorenstein \& Shapiro 1985; Schneider \& Sluse 2013; Suyu \& Alkola 2010) and the environment of the lenses (Keeton, Kochanek \& Falco 1997; Metcalf 2005). }

Very recently, Jee, Komatsu \& Suyu (2015) performed the Paraficz \& Hjorth (2009) technique on two very well-known strong lens systems, namely, B1608+656 ($z_l = 0.6304$) and RXJ1131-1231 ($z_l = 0.295$), opening a new window to observational cosmology. These authors considered a more realistic model to describe the strong lens systems by taking into account an arbitrary power-law profile (Cao et al. 2015) and  the effect of an external convergence.{These authors also explored the impact of anisotropic velocity dispersion as well as the isotropic case.}

In this paper, by considering that the results of observational cosmology in the last years have opened up an unprecedented opportunity  to investigate and compare different observational data and look for any systematic in them, we confront the $D_{Aol}$'s obtained to the B1608+656 and RXJ1131-1231 systems with luminosity distance,$D_L$, of type Ia supernovae  (SNe Ia) and   ADD of galaxy clusters obtained via their Sunyaev-Zeldovich effect (SZE) (Sunyaev \& Zeldovich 1972)  and X-ray observations. More precisely, we verify if the data sets obey  the so-called cosmic distance duality relation (Etherington 1933), $D_L(1+z)^{-2}/D_A=1$.

The paper is organized as follows: in section II we describe the method to obtain the distances to the lensing systems under study. In sec. III we show the approaches  used in our analyses and the details of the SNe Ia and galaxy clusters samples. In section IV  the results are presented and we finalize the paper in section IV with the conclusions.

\begin{figure*}
\centering
\includegraphics[width=0.45\textwidth]{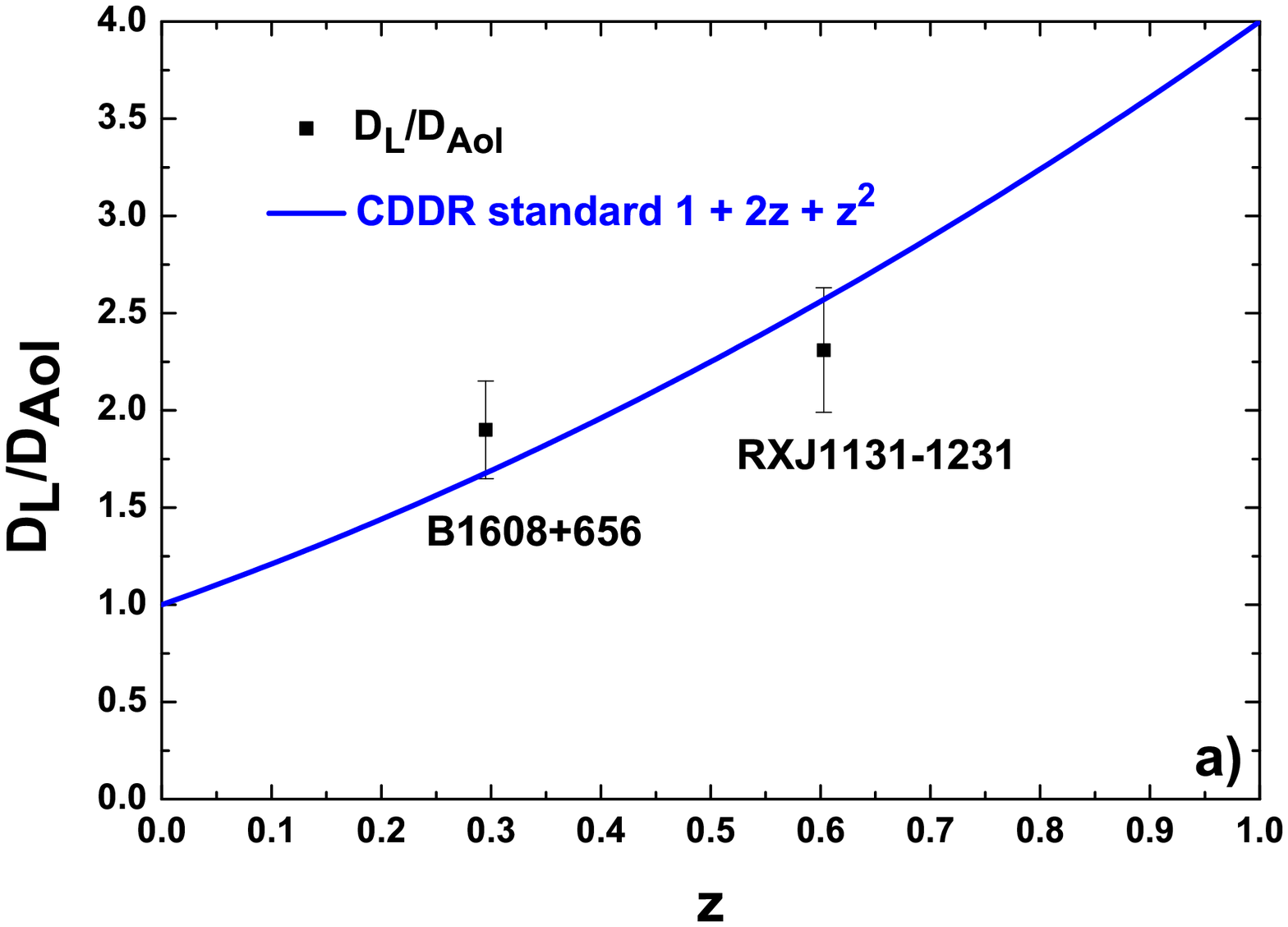}
\includegraphics[width=0.45\textwidth]{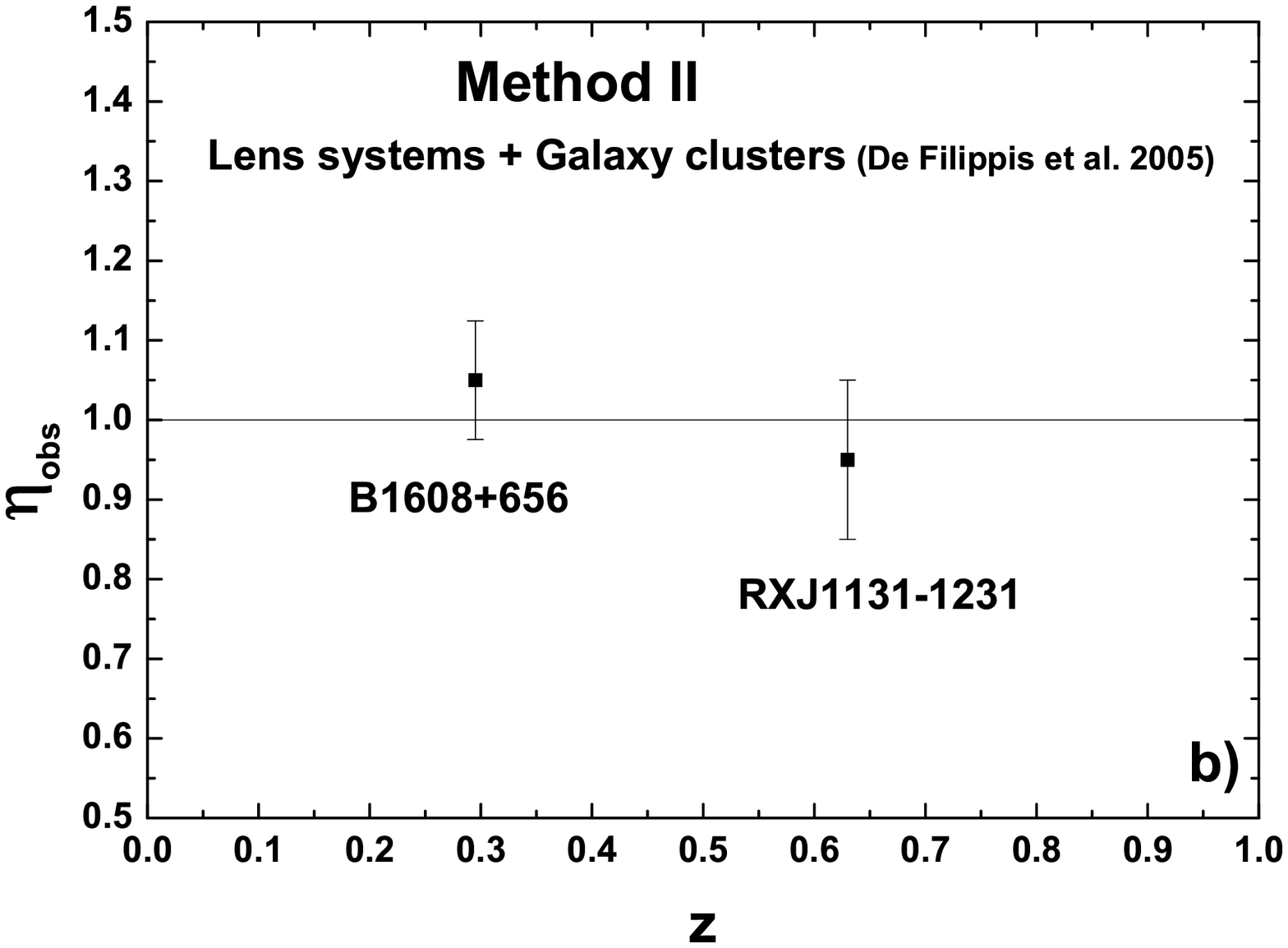}
\caption{ In fig. (a) we plot the results from the Method I. The blue solid line corresponds to standard CDDR, $D_L/D_A=1 + 2z + z^2$.  In fig. (b)  we plot the results of the analysis from method II by using De Filippis et al. (2005). In this case the standard CDDR corresponds to $\eta=1$ (horizontal black line). The compatibility between the data and the CDDR validity is verified at 1 $\sigma$ c.l.. }
\end{figure*}

\section{ $D_{Aol}$'s to the B1608+656 and RXJ1131-1231 systems}

The B1608+656 is a four-image gravitational lens system formed by a pair of interacting lens Cd galaxies with $z_l=0.6304$ and an extended
source at $z_s=1.394$ (see Fig. 2 in Jee, Komatsu \& Suyu 2015). This lens system has all three independent time delays between the images measured with errors of only a few percent, furnishing a great opportunity to measure the Hubble constant, $H_0$. For instance, by fixing the cosmological parameters Suyu et al . (2010) found $H_0 =70.6 \pm 3.1$ km/s Mpc. 

The RXJ1131-1231 system is a quadruply imaged quasar lensed with $z_s=0.6304$ and $z_l=0.295$,  discovered by Sluse et al. (2003). By using more flexible gravitational lens models with baryonic and dark matter components, Suyu et al. (2014) obtained  time-delay distance to  RXJ1131-1231 with a $6.6\%$  total uncertainty. By combining the improved time-delay distance measurements with the WMAP9 and Planck posteriors in an open $\Lambda$CDM model the curvature parameter was constrained to be $\Omega_k = 0.00^{+0.01}_{-0.02} (68\% c.l.)$. 

As commented in  Sec. I, Paraficz \& Hjorth (2009)  showed  that one can obtain the ADD between observer and lens, $D_{Aol}$, from the ratio $\Delta t/\sigma^2$, getting via  time-delay lenses a new standard ruler. The explanation is simple, as well discussed by those authors, the velocity dispersion is related to the mass through the virial theorem: $\sigma^2\propto M_{\sigma}/R$, where $M_{\sigma}$ is the mass enclosed inside the radius $R$. The  mass can be obtained by the Einstein angle $\theta_{\rm E}$ of the lensing system, $M_{\theta_{\rm E}}=\frac{c^2}{4G}  \frac{D_{Aol}D_{Aos}}{D_{Als}}\theta_{\rm E}^2$, where $R=D_{Aol}\theta_{\rm E}$. Thus, $\sigma^2 \propto \frac{D_{Aos}}{D_{Als}}\theta_{\rm E}$. Following the Eq. (1), the time-delay is proportional to $D_{Aos}D_{Aol}/D_{Als}$, in this way the ratio $\Delta t/\sigma^2$  is dependent only on the lens distance and therefore furnishes a cosmic ruler: $\Delta t/\sigma^2 \propto D_{Aol}$.  However, their analysis in simulated data was limited to the singular isothermal sphere (SIS) density profile, as well as to an isotropic velocity dispersion.

Very recently, Jee, Komatsu \& Suyu (2015) applied the Paraficz \& Hjorth (2009) technique to two real lens systems: B1608+656 and RXJ1131-1231. In order to circumvent the present difficulties in the SIS approach, these authors considered  an arbitrary power-law profile, $\rho = \rho_0 (r/r_o)^{-\gamma}$ (for SIS model $\gamma = 2$) and  the effect of an external convergence in their analyses. By considering  an isotropic velocity dispersion and spherical symmetry, the $D_{Aol}$'s obtained to the B1608+656 and RXJ1131-1231 systems were, respectively: $1485.7 \pm 208$ Mpc and $813.33 \pm 105.75$ Mpc, corresponding to a total uncertainty of $14\%$ and $13\%$. The uncertainty in the inferred distance is dominated by the velocity dispersion error, while the effect of external convergence cancels out when dividing the time delays and velocity dispersion measurements. The factor $\gamma$ of the power-law profile obtained were: $2.08 \pm 0.03$ and $1.95^{+0.05}_{-0.04}$. In the follows, since the spherical density profile  and isotropic velocity dispersion hypothesis were assumed, we present our two methods to check the robustness of the  $D_{Aol}$'s obtained to the B1608+656 and RXJ1131-1231 systems.

\section{Methods}

We verify the robustness of the $D_{Aol}$'s obtained to the B1608+656 and RXJ1131-1231 systems by using the validity of the cosmic distance duality relation (CDDR), $D_L(1+z)^{-2}/D_A=1$. This relation is valid when source and observer are connected by null geodesics in a Riemannian spacetime and the number of photons is conserved. Recently, several authors have used similar approach to search some inconsistency between other cosmological data, such as, baryon acoustic oscillations, SNe Ia, galaxy clusters, cosmic expansion rate ($H(z)$) and gravitational lenses  (Holanda, Lima \& Ribeiro 2010, 2011; Gon\c{c}alves, Holanda \& Alcaniz 2012; Chen et al. (2012); Avgoustidis et al. 2010, 2012). { As a result, the validity of the CDDR was verified at least within 2$\sigma$ c.l. (see table in Holanda, Busti \& Alcaniz 2016).}

Since three kind of astronomical observations are used in this paper,  we divide our analyses in two methods:

- Method I: we  use $D_L$'s of SNe Ia and the two $D_{Aol}$'s  to verify the consistency them  with the CDDR validity, i. e., if $D_L/D_{Aol}= (1+z)^2$. As SNe Ia data we use the Union 2.1 SNe Ia sample (Suzuki et al. 2012), an update of the original Union compilation (Amanullah et al. 2010) that comprises 580 data points  in the redshift range $0.015 < z < 1.43$. In Fig. (1a) we plot the Union 2.1 SNe Ia data.

{

However, in order to verify the CDDR with present data, we need SNe Ia with the identical redshifts to the lens systems. In our analysis, we follow the approach proposed by Meng et al. (2012) and Holanda \& Barros (2016): for each i-lens system, we obtain one  distance modulus, $\bar{\mu}$, and its error, $\sigma^2_{\bar{\mu}}$, from all i-SNe Ia  with $|z_{l_i} - z_{SNe_i}| \leq 0.005$. As stressed by those authors, this criterion allows us to have some SNe Ia for each lens system and so we can perform  a  weighted average with them in order to minimize the scatter observed on the Hubble diagram [see fig. 1(a)]. The weighted average is obtained by:

\begin{equation}
\begin{array}{l}
\bar{\mu}=\frac{\sum\left(\mu_{i}/\sigma^2_{\mu_{i}}\right)}{\sum1/\sigma^2_{\mu_{i}}} ,\\
\sigma^2_{\bar{\mu}}=\frac{1}{\sum1/\sigma^2_{\mu_{i}}}.
\end{array}\label{eq:dlsigdl}
\end{equation}
In this way, the luminosity distance, $D_L$, to each i-lens system from SNe Ia data is obtained from: $D_L(z)=10^{(\bar{\mu(z)}-25)/5}$ Mpc and   $\sigma^2_{D_L}=(\frac{\partial D_L}{\partial \mu} )^2 \sigma^2_{\bar{\mu}}$. It is important to stress that the distance moduli of Union2.1 SNe Ia compilation are dependent on the choice of the Hubble parameter $H_0=70$km/s/Mpc as well as of the $\omega$CDM cosmological model.}

- Method II: in this method, the robustness of the two $D_{Aol}$ is tested by using the Uzan et al. (2005) results. These authors argued that the SZE/X-ray technique for measuring ADD to galaxy clusters is strongly dependent on the validity of the CDDR relation, providing a test for the CDDR. When the relation does not hold,  $D_L(1+z)^{-2}/D_A=\eta$, the ADD determined from observations is $D^{c}_A=\eta^2 D_A$, where $D_A$ is the true ADD. In this way, since $D_{Aol}$ is not dependent on the CDDR validity, if one assumes $D_{Aol}=D_A$ and found $\eta =1$, reasonable assumptions about the lens systems and galaxy clusters were considered and no tension between the data will be verified. 

  We use  ADD  of galaxy clusters, $D^{c}_A$,  obtained via their Sunyaev-Zeldovich effect (Sunyaev \& Zeldovich 1980) and X-ray surface brightness observations, from a sample  compiled by De Filippis et al. (2005). This sample contains 25 galaxy clusters in low and intermediary redshifts, more precisely, $0.023< z < 0.784$. Since the standard spherical geometry has been severely questioned by the Chandra and XMMNewton
observations,  De Fillipis et al. (2005)  used an isothermal elliptical 2-dimensional $\beta$-model to describe the galaxy clusters. Previous papers that tested the validity of the CDDR with galaxy clusters observations also  found that the elliptical model is  the best geometrical hypothesis to describe these astronomical structures (Holanda, Lima \& Ribeiro 2010, 2011, 2012; Nair, Jhin-gan, \& Jain 2011; Meng, Zhang, Zhan \& Wang 2012).  The statistical contributions in galaxy clusters sample are: i) Sunyaev-Zelvovich effect  point sources $\pm 8$\%, X-ray background $\pm 2$\%, Galactic N$_{H}$ $\leq \pm 1\%$, $\pm 8$\% kinetic SZ and for CMB anisotropy $\leq \pm 2\%$.

	As for method I, it is necessary  $D_{Aol}$ and $D^{c}_A$ on the same redshift. Here, we fit the $D^{c}_A$ data with a second degree polynomial fit. For De Filippis et al. sample we used: $D^{c}_A(z)= Az + Bz^2 $, where (in Mpc), $A=3811.23784  \pm 340.66794$, $B=-2634.83172 \pm 1014.59707$ and $cov(A,B)=-298290.00426$, with $\chi^2_{red} = 1.2$. In this case, when a third degree polynomial fit is performed,
the additional term is compatible with zero. In both methods, the 1$\sigma$ error from polynomial fits is given by: 

\begin{eqnarray}
\sigma^2 & = & \left(\frac{\partial D^c_A}{\partial A}\right)^2\sigma^2_A +\left(\frac{\partial D^c_A}{\partial B}\right)^2\sigma^2_B \\ & & \nonumber + 2\left(\frac{\partial D^c_A}{\partial A}\frac{\partial D^c_A}{\partial B}\right)cov(A,B).                                                               
\end{eqnarray}

\section{Results } 

The results of our analyses are plotted in figures (2a) and (2b). In fig. (2a) we plot the results from method I. We estimate the quantity $D_L/D_{Aol}$ and obtain to B1608+656 and RXJ1131-1231, respectively: $1.9 \pm 0.25$ and $2.31 \pm 0.32$, at 1$\sigma$, in full agreement with the standard CDDR, $1 + 2z + z^2$, plotted on the blue line. 

The results from method II are plotted in figure (2b). In this case we estimate $\eta_{obs}=\sqrt{D^c_A/D_{Aol}}$. We obtain to B1608+656 and RXJ1131-1231, respectively: $\eta_{obs}=1.046 \pm 0.074$ and $\eta_{obs}=0.955 \pm 0.108$.  In both methods, the error bars are obtained simply by the trivial error propagation technique. The results are given at 1 $\sigma$ c.l. and they are in full agreement with CDDR validity. Our results show that the spherical geometry and isotropic velocity dispersion hypotheses to B1608+656 and RXJ1131-1231 systems are completely { reasonable}. It is important to comment that the CDDR was tested recently by using another quantity from strong lensing systems: the Einstein radius. Holanda, Busti \& Alcaniz (2016) used 95 galactic strong lensing systems from Sloan Lens ACS Survey (SLACS), BOSS Emission Line Lens Survey (BELLS), Lenses Structure and Dynamics Survey (LSD) and Strong Legacy Survey (SL2S) presented in Cao et al. (2015) and SNe Ia measurements. It was explored the influence of the model used to describe the lens by performing the fits under the assumptions of the singular isothermal sphere { and  with a general power-law index $\gamma$}. No violation of the CDDR  was obtained (see also Liao et al. 2016).

\section{Conclusions}

In the last twenty years, observational cosmology has become an exciting area in astronomy. Probes in several research areas has brought  a  unprecedented quantity and quality of data. For instance, the SNe Ia are limited by the systematic errors rather than statistical. So, the variety of current astronomical data provide not only the possibility of constraining cosmological parameters but also of testing some fundamental hypotheses in cosmology. However, a different route can be performed, which is to use fundamental hypotheses as guarantee in order to search some  tension between  observational data.

 In this paper we have followed this last  approach to explore a new standard ruler, the angular diameter distance (ADD) to the lens in gravitational lensing systems, which has been recently proposed.  We have taken the cosmic distance duality relation (CDDR), $D_L(1+z)^{-2}/D_A=1$, as valid and confronted the ADD to lens, $D_{Aol}$, in two strong lens systems, namely, B1608+656 and RXJ1131-1231,  with luminosity distances to SNe Ia, $D_L$,  and ADD to galaxy clusters, $D^c_A$. This is a important task given the many hypotheses  used to describe mainly the lens systems and galaxy clusters (see sections II and III). As results, we have shown that by using the two $D_{Aol}$'s and  $D_L$'s, the standard CDDR result, $D_L/D_{Aol}= 1 + 2z + z^2$ was verified at 1 $\sigma$.  When we have deformed $D_L(1+z)^{-2}/D_A=\eta$  and used the two $D_{Aol}$'s and $D^c_A$ to galaxy clusters, we also have obtained the standard relation, $\eta=1$, at 1 $\sigma$. 

In this way, our results have shown no  tension between the data and the CDDR validity. Finally, we wish to stress that in the near future, as more and larger $D_{Aol}$ data sets with smaller statistical and systematic uncertainties become available from Large Synoptic
Survey Telescope (LSST), Square Kilometre Array (SKA), Joint Dark Energy Mission (JDEM), Euclid, and the Observatory for Multi-Epoch Gravitational Lens Astrophysics (OMEGA) experiments  (Dobke et al. 2009; Ivezic et al. 2008; Carilli \& Rawlings 2004; Marshall et al. 2005; Moustakas et al. 2008), the methods proposed in this paper can improve the results on the  robustness of the $D_{Aol}$ measurements obtained via spherical geometry and isotropic velocity dispersion assumptions.

\section*{Acknowledgments}
RFLH is supported by INCT-A and CNPq (No. 478524/2013-7;303734/2014-0). The author thanks Vinicius Consolini Busti for useful comments and suggestions. The author thanks to anonymous referee by valuable suggestions.

\end{document}